\documentclass[12pt,preprint]{aastex}

\usepackage{ulem}

\def\kms  {km~s$^{-1}$}
\def\masy {mas~y$^{-1}$}

\def\etal {et al.~}

\def\meth {CH$_3$OH}

\def\Ga {G12.89+0.49}
\def\Gb {G15.03-0.68}
\def\Gc {G27.36-0.16}

\def\Vlsr {\ifmmode {V_{\rm LSR}} \else {$V_{\rm LSR}$} \fi}
\def\Ro   {\ifmmode {R_0} \else {$R_0$} \fi}
\def\To   {\ifmmode {\Theta_0} \else {$\Theta_0$} \fi}

\def\Vlsr {\ifmmode {V_{\rm LSR}} \else {$V_{\rm LSR}$} \fi}

\def\as     {\ifmmode {\rlap.}$\,$''$\,$\! \else ${\rlap.}$\,$''$\,$\!$\fi}

\slugcomment{Draft: 2011 Mar. 12}

\shorttitle{Distance to \Ga,\ \Gb\ and \Gc} \shortauthors{Xu et al.}

\begin{document}

\title{Trigonometric Parallaxes of Massive Star Forming Regions:
       VIII. \Ga,\ \Gb\ (M17)\ and \Gc}

\author{Y. Xu\altaffilmark{1}, L. Moscadelli\altaffilmark{2}, M. J. Reid\altaffilmark{3}, K. M.
Menten\altaffilmark{4}, B. Zhang\altaffilmark{5}, X. W.
Zheng\altaffilmark{6}, A. Brunthaler\altaffilmark{3}}

\altaffiltext{1}{Purple Mountain Observatory, Chinese Academy of
Sciences, Nanjing 210008, China; xuye@pmo.ac.cn}
\altaffiltext{2}{Osservatorio di Arcetri, Firenze,
Italy}\altaffiltext{3}{Harvard-Smithsonian Center for Astrophysics,
60 Garden Street, Cambridge, MA 02138,
USA}\altaffiltext{4}{Max-Planck-Institut f$\ddot{u}$r
Radioastronomie, Auf dem H{\" u}gel 69, 53121 Bonn, Germany}
 \altaffiltext{5}{Shanghai Observatory, Chinese
Academy of Sciences, Shanghai 200030, China}\altaffiltext{6}{Nanjing
University, Nanjing 20093, China}

\begin{abstract}
We report trigonometric parallaxes for three massive star forming
regions, corresponding to distances of \ $2.34^{+0.13}_{-0.11}$~kpc
for \Ga\ (also known as IRAS 18089$-$1732),\
$1.98^{+0.14}_{-0.12}$~kpc for \Gb\ (in the M17 region),
and $8.0^{+4.0}_{-2.0}$~kpc for
\Gc.  Both \Ga\ and \Gb\ are located in the Carina-Sagittarius
spiral arm.

\end{abstract}

\section{Introduction}
Accurately determining the spiral structure of the Milky Way is a
difficult task. In principle, kinematic distances can be used to
construct the spiral structure of the Galaxy. However, recent work
on parallax and proper motion measurements has shown that kinematic
distances can be affected by large uncertainties. In some places of
the Galaxy, characterized by ``anomalous'' motions (as, for example,
the Perseus arm \citep{Xu:06a}), kinematic distances can be in error
by a factor as large as two. Relying solely on kinematic distances,
one cannot accurately determine the location of spiral arms. A more
secure method of distance measurement is required to re-construct
the Galactic spiral structure and its 3-D motion \citep{Reid:09a}.
Recently, \meth \ 12~GHz masers have been used as astrometric
targets to measure the trigonometric parallaxes and proper motions
of massive star-forming regions \citep[][hereafter called
Paper~I]{Reid:09b}. As part of that large project, here, we present
the results of our parallax measurement campaign toward the sources
\Ga,\ \Gb\ and \Gc.

\section{Observations and Calibration}

A series of  observations of the 12~GHz \meth\ masers
in the \Ga, \Gb, and \Gc\ star-forming regions
were carried out with the NRAO\footnote{The National Radio Astronomy
Observatory is a facility of the National Science Foundation
operated under cooperative agreement by Associated Universities,
Inc.} Very Long Baseline Array (VLBA).  Paper~I provides
a description of  the general
observational setup and data calibration procedures, and
here we note only details specific to the observations of the three
sources presented in this paper.

\Ga\ and \Gb\ were observed at five epochs (VLBA program BR129A):
2007 October 18; April 17 and  September 13 2008; March 23 and
October 23 2009. However, because of bad weather, the data quality
of the first epoch was not good enough for parallax measurement, and
the present results are based on the following four epochs only.
\Gc\ was observed at four epochs (VLBA program BR129C): 2007 October
27; April 24 and October 31 2008; 2009 April 16. The observing dates
have been chosen to sample the peaks of the parallax signature in
right ascension, as the amplitude of the parallax signature in
declination is considerably smaller.

For each maser source, we used three different background sources,
selected from the following calibrator surveys: the VLBI Exploration
of Radio Astrometry (VERA, \citet{Honma:01}) Galactic Plane Survey;
a VLA survey of compact NVSS sources \citep{Xu:06b}; the VCS2 and
VCS3 catalogs \citep{Fomalont:03, Petrov:05}. Only background
sources belonging to the VCS2 and VCS3 catalogs were detected. The
non-detected calibrators were J1815-1717 and J1837-0628 from the
VERA survey, and J1809-1804 and J1817-1614 from the VLA survey.
Table~\ref{table:positions} reports the peak positions and
intensities of the maser and background sources, listing also the
main observing parameters.  We recorded four adjacent dual
circularly polarized 4~MHz bands with the maser signal in the second
band centered at the peak velocity of the 12~GHz maser emission,
corresponding to a LSR velocity, \Vlsr, of 40~\kms, 23~\kms\ and
100~\kms\ for source \Ga, \Gb\ and \Gc,\ respectively. The spectral
resolution was 0.38~\kms.

We used observations of the strong VLBA calibrators J1800$+$3848 and
3C345 (J1642+3948) to correct for instrumental delays and phase
offsets among different frequency bands. The spectral channel with
the strongest maser emission was used as the phase reference (see
Table 1). The maser reference features were detected at all epochs
and were relatively stable in flux, with the exception of the source
\Ga, whose correlated flux density on the shortest VLBA baselines
increased monotonically by a factor of $\approx$ 2, while
\citet{Goedhart:09} found a periodical variation with a period of
$\sim$30 days and flux variations of $\sim$50\%. After applying the
maser calibration, we integrated the data of the background
continuum sources over all four dual-polarized bands and imaged the
sources using the AIPS task IMAGR. For all maser targets, the
maser-referenced calibrator image has a dynamical range of about 10.

The naturally-weighted ``dirty'' beam was determined by the
availability of maser phase-reference data and, as expected for
low-declination sources, was strongly elongated close the N--S
direction. For each of the three maser targets,
Table~\ref{table:positions} reports the parameters of the ``dirty''
beam as derived by the AIPS task IMAGR by fitting an elliptical
Gaussian brightness distribution to the central portion of the
``dirty'' beam map. The images of the maser and corresponding
calibrators  have been restored using a round "clean" beam, with a
FWHM size  intermediate between the minor and major FWHM sizes of
the ``dirty'' beam. To make images from different epochs more
readily comparable, maps of a given source were restored using the
same beam.

After obtaining maps of both maser and background sources for all
epochs, we derived the parameters of the emission by fitting
elliptical Gaussian brightness distributions.
Table~\ref{table:positions} lists the position and peak intensity of the
maser and corresponding calibrator(s),  as well as the angular
separation on the sky between the two sources. Absolute maser
positions are derived from the positions of the ICRF sources
J1825$-$1718 and J1834$-$0301, which are accurate within \
$\approx1$~mas.

\subsection{Methanol masers}
Emission in the 12.2 GHz \meth\  maser line generally is coextensive
with emission in the 6.7 GHz line -- maser spots at similar
velocities often arise from similar positions \citep{Menten:92}.
Since the 6.7 GHz emission is almost always (much) stronger than the
12.2 GHz emission, spectra in the former are much richer and more
maser spots are detected at 6.7 GHz \citep[see, e.g.
][]{Caswell1995}. For all our sources, the emission is  dominated by
a single  feature at the LSR velocity listed in
Table~\ref{table:positions}. The positions of any other weak
features (if present) were found to be close (withih a few mas) to
the strongest feature.

\section{Results}

Following the analysis described in Paper~I, the measured positions
of the masers were modeled as a linear combination of the elliptical
parallax and linear proper motion signatures. Because systematic
errors (owing to small uncompensated atmospheric delays and, in some
cases, varying maser and calibrator source structures) typically
dominate over thermal noise when measuring relative source
positions, we added ``error floors'' in quadrature to the formal
position uncertainties. We used different error floors for the Right
Ascension and Declination data and adjusted them to yield post-fit
residuals with $\chi^2$ per degree of freedom near unity for both
coordinates.

\subsection{\Ga} \label{PPM_g12}

\begin{figure}
\includegraphics[angle=-90,scale=0.5]{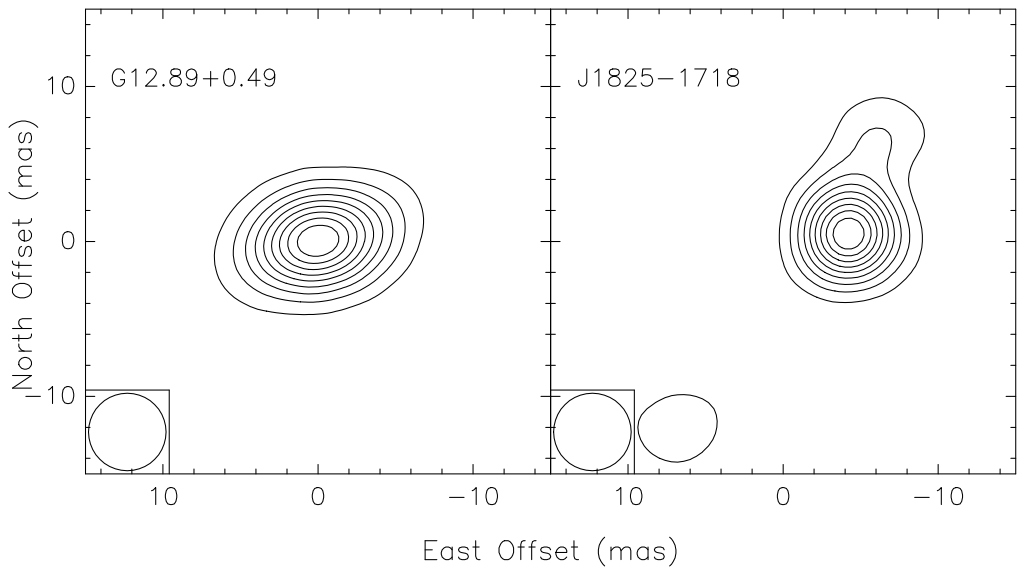}
\caption{The {\it left} and {\it right} panels present the map of
 the \Ga\ reference maser channel (\Vlsr = 39.8~\kms)
 and the background source J1825$-$1718, phase-referenced
to the reference maser channel, respectively.
Both maps are for the epoch 2008 April 17.
Contour levels are at integer multiples (with the zero contour suppressed)
of 10\% of the peak brightness of
2.2~Jy~beam$^{-1}$ for \Ga\ and 0.14~Jy~beam$^{-1}$ for
J1825$-$1718. The FWHM size of the restoring beam is given in the
lower left corner of each panel. \label{g12_masers}}
\end{figure}

\begin{figure}
\includegraphics[angle=-90,scale=0.67]{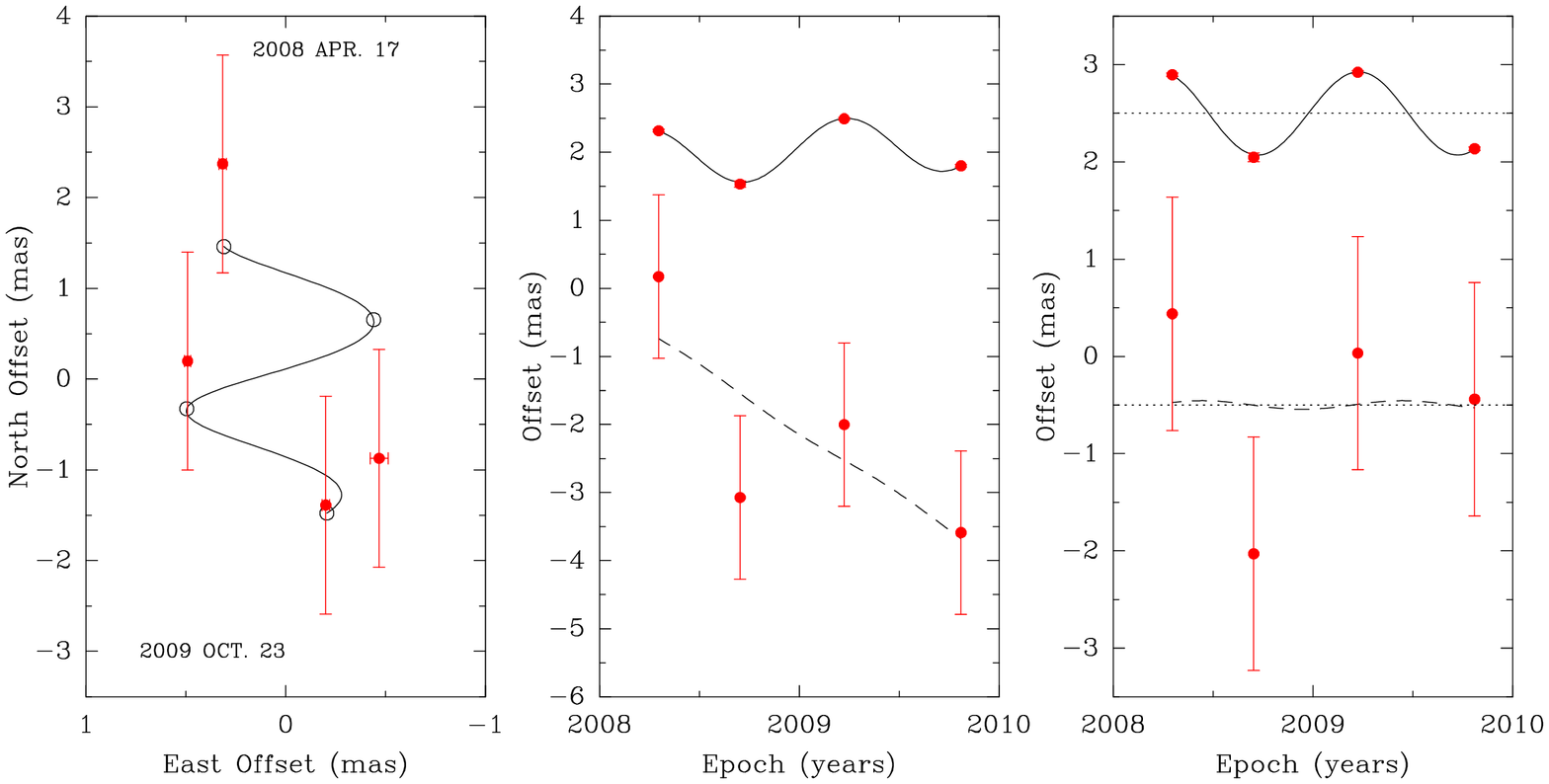}
\caption{Parallax and proper motion data and fits for \Ga. Plotted
are position offsets of the \Ga\ maser spot at \Vlsr = 39.8~\kms\
relative to the background source  J1825$-$1718. {\it Left
Panel:} Positions on the sky with first and last epochs labeled. The
expected positions from the parallax and proper motion fit are
indicated {\it (empty circles)}.
{\it Middle Panel:} The position offsets of the maser along the East
and North direction versus time.  The best-fit model of the variation of the
East and North offsets with time is shown as
{\it continuous} and {\it dashed} lines, respectively.
{\it Right Panel:}
Same as the {\it middle panel}, except the best fit proper motions
have been removed, allowing the effects
of only the parallax to be seen.
\label{g12_parallax}}
\end{figure}

Fig.~\ref{g12_masers} presents the images of the \Ga\ reference
maser channel at \Vlsr = 39.8~\kms\ and the background continuum
source J1825$-$1718 (phase referenced to the reference maser
channel) at the second epoch (2008 April 17). Both the maser and the
background source emission is dominated by a single compact
component, which serves as an astrometric target.
Fig.~\ref{g12_parallax} reports the positions of the emission in the
reference maser channel (relative to the background source
J1825$-$1718) as a function of time and the parallax fit. The large
declination errors of the parallax fit are probably caused by
residual phase errors due to the combination of the large
maser-calibrator separation (3\fdg3) and the low maser declination.

Fitting for the parallax and proper motion simultaneously, we obtain
$\pi=0.428 \pm 0.022$~mas. The proper motions in the eastward and
northward directions are $0.16\pm0.03$ and $-1.90
\pm1.59$~mas~y$^{-1}$, respectively (see Table~\ref{table:allfits}).
This parallax corresponds to a distance of
$2.34^{+0.13}_{-0.11}$~kpc. Based on this distance, \Ga\ is most
likely in the Carina-Sagittarius arm of the Milky Way, and not in
the Crux-Scutum arm.

\subsubsection{Maser environment}
\Ga, better known as IRAS~18089$-$1732, has long be known as a
prominent maser source, showing strong emission from all the
widespread interstellar maser molecules (OH, H$_2$O, and \meth). It
was included in the sample of high mass protostellar objects (HMPOs)
defined by \citet{Sridharan:02} and further studied by \citet[][ and
references therein]{Beuther:02a}. Except for a single spot (the
component~F) detached by 1$\arcsec$.4, all of the 6.7 GHz \meth\
maser spots detected by \citet{Walsh:98} coincide with our VLBA
maser position within the errors ($\sim$1$\arcsec$) of the Australia
Telescope Compact Array (ATCA). Our VLBA maser emission is most
likely corresponding with the component~A of \citet{Walsh:98}, which
shows a similar \Vlsr = 39.2~\kms. H$_2$O maser emission has been
found with the VLA $\approx 1\as2$ to the NNE \citep{Beuther:02b}.
Interestingly, the H$_2$O maser position coincides with weak
cm-wavelength radio continuum emission and a (sub)mm dust emission
peak, while the \meth\ maser is offset from these peaks
\citep{Beuther2004}.  The masers and continuum emission are
associated with a rich molecular hot core, which has been
interpreted as a rotating disk \citep{Beuther2005}.

\subsection{\Gb} \label{PPM_G15}
\begin{figure}
\includegraphics[angle=-90,scale=0.5]{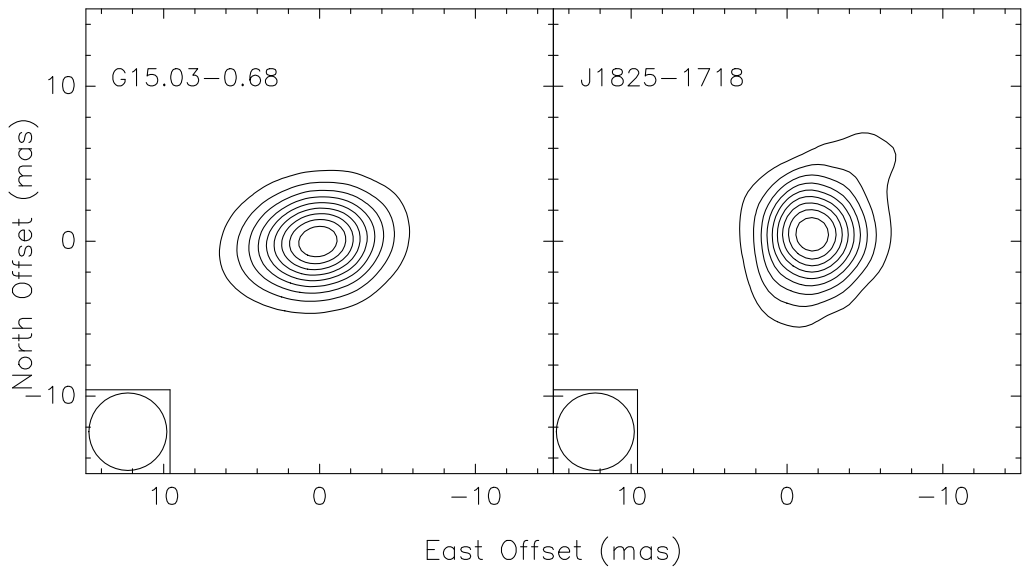}
\caption{The {\it left} and {\it right} panels present maps of the
\Gb\ reference maser channel (\Vlsr = 23.4~\kms) and the background
source J1825$-$1718, phase-referenced to the reference maser
channel, respectively. Both maps are for 2008 April 17. Contour
levels are integer multiples (with zero contours suppressed) of 10\%
of the peak brightness of 2.9~Jy~beam$^{-1}$ for \Gb\ and
0.14~Jy~beam$^{-1}$ for J1825$-$1718. The FWHM size of the restoring
beams is given in the lower left corner of each panel.
\label{g15_masers}}
\end{figure}

\begin{figure}
\includegraphics[angle=-90,scale=0.67]{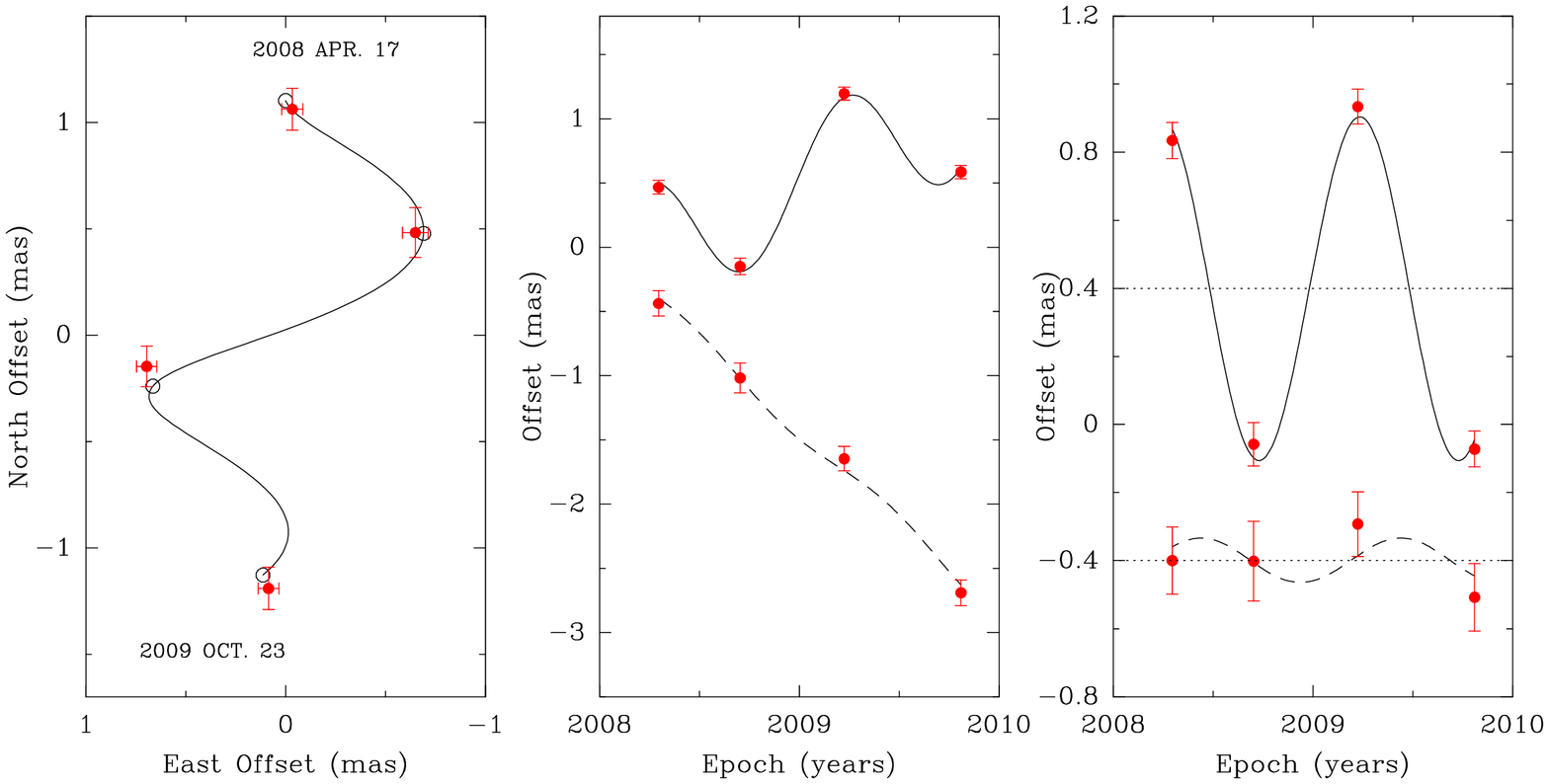}
\caption{Parallax and proper motion data and fits for \Gb. Plotted
are position offsets of the \Gb\ maser spot at \Vlsr = 23.4~\kms\
relative to the background source \ J1825$-$1718. {\it Left
Panel:} Positions on the sky with first and last epochs labeled. The
expected positions from the parallax and proper motion fit are
indicated {\it (empty circles)}.
{\it Middle Panel:} The position offsets of the maser along the East
and North direction versus time.  The best-fit model of the variation of the
East and North offsets with time is shown as
{\it continuous} and {\it dashed} lines, respectively.
{\it Right Panel:}
Same as the {\it middle panel}, except the best fit proper motions
have been removed, allowing the effects
of only the parallax to be seen.
\label{g15_parallax}}
\end{figure}

Fig.~\ref{g15_masers} presents the images of the \Gb\ reference
maser channel (\Vlsr = 23.4~\kms) and the background continuum
source J1825$-$1718 (phase referenced to the reference maser
channel), at the second epoch (2008 April 17).
Fig.~\ref{g15_parallax} reports the positions of emission in the
reference maser channel  (relative to the background source
J1825$-$1718) as a function of time and the parallax fit.

Fitting for the parallax and proper motion simultaneously, we obtain
$\pi = 0.505\pm0.033$~mas, corresponding to a distance of \
$1.98^{+0.14}_{-0.12}$~kpc. At this distance, \Gb\ is likely in the
Carina-Sagittarius arm. The proper motions in the eastward and
northward directions are $0.68\pm0.05$ and \ $-1.42
\pm0.09$~mas~y$^{-1}$, respectively (see Table~\ref{table:allfits}).

\subsubsection{Maser environment}
The maser position coincides with the ``unusual radio point source
in M17'' found by \citet{Felli:80}, one of the first ultracompact
HII regions ever identified as such. The position of the 6.7 GHz
maser emission, as recently determined by \citet{Caswell:09} using
ATCA, is consistent within the ATCA measurement error ($\approx
1''$) with that we determine for the 12.2 GHz line.

\subsection{\Gc} \label{PPM_G27}
\begin{figure}
\includegraphics[angle=-90,scale=0.5]{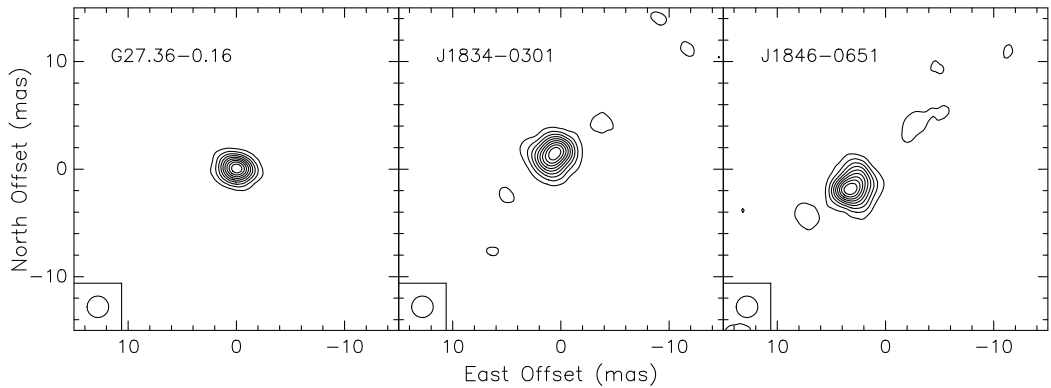}
\caption{The panels from {\it left} to {\it right} present the map
of the \Gc\ reference maser channel (\Vlsr = 99.6~\kms) and
the background sources J1834$-$0301 and \ J1846$-$0651,
phase-referenced to the reference maser channel.
The maps are for 2008 April 24.
Contour levels are integer multiples (with the zero contour suppressed)
of 10\% of the peak brightness of 2.6~Jy~beam$^{-1}$ for \Gc, and
0.07~and~0.04~Jy~beam$^{-1}$ for J1834$-$0301 and \ J1846$-$0651,
respectively. The FWHM size of the restoring beams is given in the
lower left corner of each panel. \label{g27_masers}}
\end{figure}

\begin{figure}
\includegraphics[angle=-90,scale=0.67]{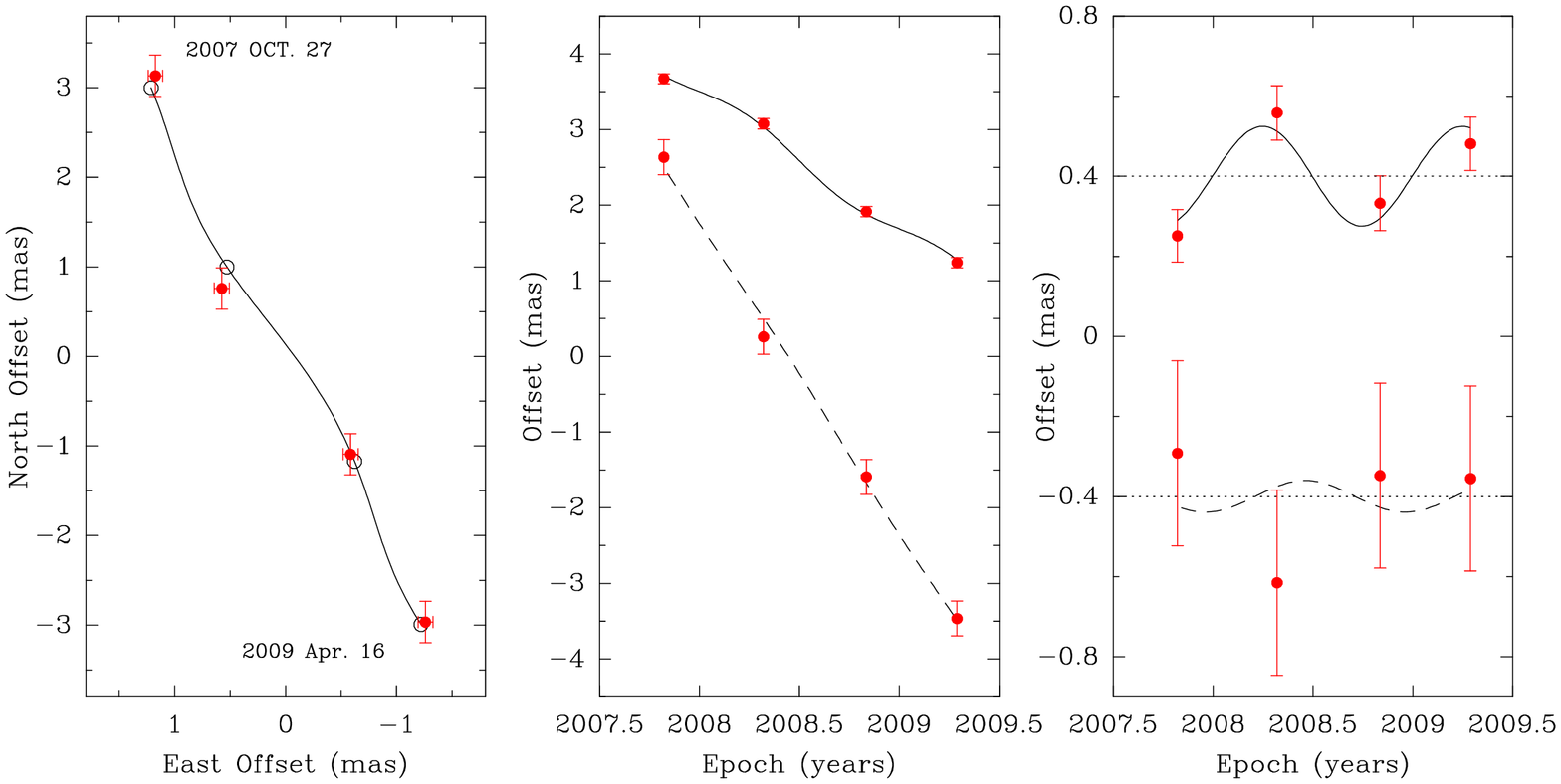}
\caption{Parallax and proper motion data and fits for \Gc. Plotted
are position offsets of the \Gc\ maser spot at \ \Vlsr = 99.6~\kms
relative to the background source \ J1834$-$0301. {\it Left Panel:}
Positions on the sky with first and last epochs labeled. The
expected positions from the parallax and proper motion fit are
indicated {\it (empty circles)}.
{\it Middle Panel:} The position offsets of the maser along the East
and North direction versus time.  The best-fit model of the variation of the
East and North offsets with time is shown as
{\it continuous} and {\it dashed} lines, respectively.
{\it Right Panel:}
Same as the {\it middle panel}, except for the fact that the best fit proper motions
have been removed, allowing the effects
of only the parallax to be seen. \label{g27_parallax}}
\end{figure}

Fig.~\ref{g27_masers} presents the images of the \Gc\ reference maser
channel (\Vlsr = 99.6~\kms) and the two background continuum sources
J1834$-$0301 and J1846$-$0651 (phase referenced to the reference maser channel)
at the second epoch (2008 April 24). Fig.~\ref{g27_parallax} reports
the positions of the reference maser channel (relative to the
background sources \ J1834$-$0301) as a function of time and the
parallax fit.

For this source, the derived value of parallax has a large
fractional uncertainty. Using only data from the \ J1834$-$0301 \
calibrator, we obtain \  $\pi = 0.125\pm0.042$~mas, while using the
\ J1846$-$0651 \ data, we find \ $\pi = 0.198\pm0.133$~mas.
Since the result from the \ J1834$-$0301 \ calibrator appears
significantly more precise than that obtained from
J1846$-$0651, we take the former as the best parallax
measurement for \Gc, corresponding to a distance of
$8.0^{+4.0}_{-2.0}$~kpc. The proper motions in the eastward and
northward directions are $-1.81\pm0.08$ and
$-4.11\pm0.26$~mas~y$^{-1}$, respectively (see
Table~\ref{table:allfits}).

\subsubsection{Maser environment}
The 12.2 GHz \meth\ maser in the \Gc\ region is offset by about
$2''$ from a compact $< 2''$ radio source found by
\citet{Becker:94}, which also has an associated IRAS source
(18391$-$0504). Much stronger emission than in this line has been
found in the 6.7 GHz \meth\  line \citep{Szymczak:02}, which has associated maser
emission from OH and H$_2$O  \citep{Szymczak:04,Szymczak:05}. This
source appears to have been little studied in thermal molecular line
emission.

\section{Galactic Locations and 3-D Motions}

Combining the distances, LSR velocities and proper motions of the
masers yields their locations in the Galaxy and their full space
motions. Since internal motions of 12~GHz methanol masers are fairly
small, typically $\sim3$~\kms \citep{Mos02}, the maser motions
should be close to that of their associated young stars.  Given a
model for the scale and rotation of the Milky Way, we can subtract
the effects of Galactic rotation and the peculiar motion of the Sun
from the space motions of the maser sources and estimate the
peculiar motions of the associated young stars. Following the
discussion by \citet{Reid:09a}, the motion of an individual massive
star (associated with the masers) can then be taken as
representative of that of the whole star-forming region to which the
massive star belongs, allowing for a velocity dispersion of
individual stars of $\approx 7$~\kms\ per each velocity coordinate.
Peculiar motions are given in a Galactocentric reference frame,
where  $U, V \, {\rm and} \, W$  are the velocity components toward
the Galactic center, in the direction of Galactic rotation, and
toward the North Galactic Pole, respectively, at the location of a
given source in the Galaxy. We adopt the IAU values for the distance
to the Galactic center ($R_0=8.5$~kpc) and the rotation speed of the
Galaxy at this distance ($\Theta_0=220$~\kms), and assumes a flat
rotation curve. We use the Solar Motion value ($U = 11.1
^{+0.69}_{-0.75}$, $V = 12.24 ^{+0.47}_{-0.47}$, and $W = 7.25
^{+0.37}_{-0.36}$~\kms) by \citet{Schonrich:10}, which have recently
revised the Hipparcos satellite result. Table~\ref{table:3dmotion}
reports the peculiar motions derived for our maser targets. For
comparison, the results obtained with a different model of Galactic
rotation ($R_0 = 8.3 \pm 0.23$~kpc, $\Theta_0 = 239 \pm 7 $~\kms),
based on both the weighted average of recent four direct
measurements of the distance to the Galactic center and the proper
motion of Sgr A$^{\ast}$ \citep{Brunthale2011}, are also reported.
Looking at Table~\ref{table:3dmotion}, one can see that, except for
\Gc\, whose space motion may be strongly affected by the large
parallax uncertainty, the derived peculiar motions are fairly
independent on the adopted Galactic rotation model. For all sources,
the motion directed toward the north Galactic pole is  only of a few
\kms. This is consistent with the expectation for the motion of
massive star-forming regions, which should be mainly in the Galactic
plane.

\section{Conclusions}
We have measured the parallax and proper motion of 12~GHz methanol
masers in three star-forming regions. For two sources, the derived
distances are accurate by better than 10\%:
$2.34^{+0.13}_{-0.11}$~kpc for \Ga\ and $1.98^{+0.14}_{-0.12}$~kpc
for \Gb. For the source \Gc, the derived distance is affected by a large
uncertainty: $8.0^{+4.0}_{-2.0}$.

Our precise absolute positions place the methanol masers near the center
of active regions of high mass star formation, as traced by molecular hot cores,
ultracompact HII regions and/or dust condensations or a combination of these.

\acknowledgments This work was supported by the Chinese NSF through
grants NSF 11073054, NSF 10733030, NSF 10703010 and NSF 10621303,
and NBRPC (973 Program) under grant 2007CB815403.

\vskip 0.5truecm
{\it Facilities:} \facility{VLBA}

\begin{deluxetable}{lllllll}
\tablecolumns{6} \tablewidth{0pc} \tablecaption{Positions and
Brightnesses} \tablehead{ \colhead{Source} & \colhead{R.A. (J2000)} &
\colhead{Dec. (J2000)} & \colhead{$\phi$} & \colhead{Brightness} &
\colhead{\Vlsr}
& \colhead{NW beam} \\
 \colhead{} & \colhead{$\mathrm{(^h\;\;\;^m\;\;\;^s)}$}
& \colhead{$(\degr\;\;\;\arcmin\;\;\;\arcsec)$} &
\colhead{($^{\circ}$)} & \colhead{(Jy/beam)} & \colhead{(\kms)}&
\colhead{(mas, mas, deg)} } \startdata
 \Ga          & 18~11~51.3955   & $-$17~31~29.913   &      & 2.2  & 39.8 &  4.2$\times$2.5 @ 2   \\
 J1825-1718   & 18~25~36.53228  & $-$17~18~49.8484  & 3.3  & 0.14 &      &  5.7$\times$3.8 @ $-$33 \\
 \\
 \Gb          & 18~20~24.8111   & $-$16~11~35.316   &      & 2.9  & 23.4 &  3.6$\times$1.8 @ $-$1  \\
 J1825-1718   & 18~25~36.53228  & $-$17~18~49.8484  & 1.7  & 0.14 &      &  4.3$\times$2.2 @ $-$11 \\
 \\
 \Gc          & 18~41~51.0570   & $-$05~01~43.443   &      & 2.6  & 99.6 &  2.7$\times$1.4 @ 16 \\
 J1834-0301   & 18~34~14.07456  & $-$03~01~19.6274  & 2.8  & 0.07 &  & 2.8$\times$1.7 @ 13   \\
 J1846-0651   & 18~46~06.30026  & $-$06~51~27.7456  & 2.1  & 0.04 &  & 3.1$\times$1.7 @ 20   \\
\enddata
\tablecomments {$\phi$ is the angular separation between the maser
and the calibrator. The maser absolute position, the peak
brightness,  the size and P.A. of the naturally-weighted (NW) beam
are listed for the epoch \ 2008 April 17 \ and \ 2008 April 24, for
the source couple \Ga\ and \Gb, and \Gc,\ respectively. The P.A. of
the beam is defined as East of North.} \label{table:positions}
\end{deluxetable}

\begin{deluxetable}{lllllll}
\tablecolumns{5} \tablewidth{0pc} \tablecaption{ Parallax \& Proper
Motion Fit}
\tablehead {
 \colhead{Maser} & \colhead{ \Vlsr} & \colhead{Background} &
  \colhead{Parallax} & \colhead{$\mu_x$} &
  \colhead{$\mu_y$}
\\
  \colhead{Name}      &  \colhead{(\kms)}      & \colhead{Source} &
  \colhead{(mas)} & \colhead{(\masy)} &
  \colhead{(\masy)}
            }
\startdata
G12.89$+$0.49 & 39.8 & J1825-1718 & 0.428$\pm$0.022 &   0.16$\pm$0.03  & $-$1.90$\pm$1.59 \\
G15.03$-$0.68 & 23.4 & J1825-1718 & 0.505$\pm$0.033 &   0.68$\pm$0.05  & $-$1.42$\pm$0.09 \\
G27.36$-$0.16 & 92.2 & J1834-0301 & 0.125$\pm$0.042 &$-$1.81$\pm$0.08  & $-$4.11$\pm$0.26 \\
              & 92.2 & J1846-0651 & 0.198$\pm$0.133 &$-$1.90$\pm$0.27  & $-$4.75$\pm$0.15 \\
\enddata
\label{table:allfits}
\end{deluxetable}

\begin{deluxetable}{lllllllll}
\tablecolumns{8} \tablewidth{0pc} \tablecaption{Peculiar Motions}
\tablehead {
 \colhead{} & \multicolumn{3}{c}{R$_{0}$ = 8.5 kpc, $\Theta_{0}$ = 220 km s$^{-1}$} &
 & \multicolumn{3}{c}{R$_{0}$ = 8.3 kpc, $\Theta_{0}$ = 239 km s$^{-1}$}
\\
  \colhead{Maser} &  \colhead{U}  & \colhead{V} &
  \colhead{W} & & \colhead{U} & \colhead{V} & \colhead{W}
\\
  \colhead{Name} & \colhead{(km s$^{-1}$)} & \colhead{(km s$^{-1}$)} &
  \colhead{(km s$^{-1}$)} & & \colhead{(km s$^{-1}$)} &
  \colhead{(km s$^{-1}$)} & \colhead{(km s$^{-1}$)}}
\startdata
\Ga\ &  +16$\pm$7    &  +2$\pm$15  & $-$4$\pm$8 & &   +14$\pm$7   &  +2$\pm$15   & $-$4$\pm$8  \\
\Gb\ &   +2$\pm$5    &  +6$\pm$2  & $-$5$\pm$1 & &    +0$\pm$5   &  +6$\pm$2   & $-$5$\pm$1  \\
\Gc\ &  $-$53$\pm$43 & $-$32$\pm$17 & $-$3$\pm$6 & & $-$80$\pm$47  & $-$48$\pm$25  & $-$3$\pm$6  \\
\enddata
\tablecomments {For the source \Gc, the reported peculiar motion is calculated from the proper motion values
derived using data from the calibrator J1834$-$0301.}
\label{table:3dmotion}
\end{deluxetable}

\end{document}